\documentclass{PoS}
\usepackage{amsmath}
\title{Breaking of de Sitter Symmetry}

\ShortTitle{Breaking  of de Sitter Symmetry}

\author{\speaker{Myron Bander}\footnote{UCI-TR-2010-29}\\
        Department of Physics and Astronomy,\\University of California, Irvine, California 92697, USA\\
        E-mail: \email{mbander@uci.edu}}

\abstract{We show that an interacting spin-0 field on a de Sitter space background will break the underlying de Sitter symmetry.  This is done first for a (1+1) de Sitter space where a boson-fermion correspondence permits us to solve certain interacting theories by transforming them into free ones of opposite statistics.  A massless boson interacting by a sine-Gordon potential is shown to be equivalent to a free massive fermion with the mass depending on the de Sitter time thus breaking the symmetry explicitly.  We then show that for larger dimensions and any boson potential, to one loop, an anomaly develops and the currents generating the de Sitter transformations are not conserved. }

\FullConference{Quarks, Strings and the Cosmos - H\'{e}ctor Rubinstein Memorial Symposium\\
		August 09-11, 2010\\
		AlbaNova )Stockholm) Sweden}

\begin{document}

\hfill{\it To Hector, friend and colleague for fifty two years.}

\section{Introduction}\label{introduction}
De Sitter space is fundamental to our understanding of cosmology.  This geometry is a solution to Einstein's equations whose driving term is a cosmological constant; it represents the past universe during an inflationary period and also represents the future of the current accelerating scenario. However, quantum field theory on such a de Sitter background presents problems, primarily in the choice of vacuum on which to build the theory \cite{Mottola:1984ar,Allen:1985ux}. The least problematic vacuum and the one used most often is the Euclidian or Bunch-Davies vacuum \cite{Bunch:1978yq}; the propagator in this vacuum is an analytic continuation of the correlator on a sphere and has no singularities besides the one at coincident points.  All other vacua, referred to as $\alpha$-vacua \cite{Chernikov}, have an additional singularity in their propagator and, as a result, have serious difficulties with analiticity of scattering amplitudes \cite{Einhorn-Larsen} or with unitarity \cite{Akhmedov:2008pu}. Nevertheless, Polyakov \cite{Polyakov:2007mm, Polyakov:2009nq} has argued that the criterion for which vacuum to use should be based on the behavior of the propagator at large geodesic distances.  With $l$ the geodesic distance, the propagator for a particle of mass $m$ should vary as  $\exp(-iml)$  rather than a sum of $\exp(iml)$ and $\exp(-iml)$; the latter is the behavior in all vacua save the one advocated in \cite{Polyakov:2007mm, Polyakov:2009nq}. In this vacuum, as well as all the other $\alpha$ vacua, the propagator has an additional ``infrared" singularity at the antipodal point. Consequences of using such propagators for interacting field theories are quite dramatic. In \cite{Polyakov:2009nq} it is shown that to order $g^2$ the vacuum energy for a massive scalar field with a $g\phi^4$ interaction develops an imaginary part proportional
to the space-time volume.  This maybe interpreted as an explosive matter production canceling the the curvature of the underlying space. By screening this curvature de Sitter symmetry is broken. This would have severe consequences for the aforementioned use of de Sitter space in cosmology. This picture was confirmed in the case of a two dimensional, (1+1), space \cite{Bander:2010pn} where by the use of a fermion-boson correspondence certain interacting theories can be solved exactly. It was found that a massless field with a sine-Gordon interaction corresponds to a free fermion one with a de Sitter time dependent mass, explicitly breaking de Sitter symmetry. A subsequent study \cite{Bander-2} of the conservation of currents generating de Sitter symmetries showed that for interacting scalar field theories, due to the aforementioned antipodal, infrared singularity,  an anomaly develops and these currents are not conserved.

In the next section, Sec.~\ref{dSspace}, we shall review parametrization of de Sitter space and its isometries. The two dimensional, $(1+1)$, soluble field theories are discussed in Sec.~\ref{2-d}. This is a summary of \cite{Bander:2010pn} The general non conservation of de Sitter currents for interacting scalar theories is discussed in Sec.~\ref{Anomaly}; technical details promised in \cite{Bander-2} are presented here. Comments and conclusions are made in Sec.~\ref{conclusions}

\section{de Sitter Space}\label{dSspace}
A D-dimensional de Sitter space, with coordinates $\tau, x_1,\cdots, x_d$, with $d=D-1$, may be imbedded in a flat (D+1) Minkowski space with coordinates $Y_0, Y_1,\cdots, Y_D$, satisfying the constraint
\begin{equation}\label{imbed}
Y_0^2-Y_1^2-\cdots -Y_D^2=-1\,  ;
\end{equation}
the Hubble parameter is set to one. The parametrization we shall use is the flat slicing one \cite{Spradlin:2001pw} with conformal time where the metric of the space is
\begin{equation}\label{metric}
ds^2=d\tau^2-e^{2\tau}d{\vec x}\cdot d{\vec x}\, .
\end{equation}
The relation between the intrinsic de Sitter coordinates $\tau, {\vec x}$ and the embedding ones $Y_0,{\vec Y},Y_D$ (${\vec Y}$ denotes a $d$ dimensional vector) are
\begin{eqnarray}\label{conformalparametrization}
Y_0&=&\frac{1}{2}\left(\tau-\frac{1}{\tau}-\frac{{\vec x}^2}{\tau}\right)\, ,\nonumber\\
Y_i&=&-\frac{x_i}{\tau}\ \ \ (i=1\cdots d)\, ,\\
Y_D&=&\frac{1}{2}\left(-\tau-\frac{1}{\tau}+\frac{{\vec x}^2}{\tau}\right)\, .\nonumber
\end{eqnarray}
There is a one-to-one correspondence between the imbedding coordinates $Y$, satisfying (\ref{imbed}) and the intrinsic de Sitter coordinates $\tau,{\vec x}$. With $Y$ corresponding to $x=(\tau,{\vec x})$ we shall have use of the antipode ${\bar x}$ corresponding to the parity-time reversed $Y$, namely $-Y$. As mentioned in Sec.~\ref{introduction} some propagators have singularities when one point is coincident with the antipode of the other point.
We shall be interested in how the isometries of de Sitter space are implemented in the metric of (\ref{metric}). In the embedding space these isometries are Lorentz transformations involving $Y_0,{\vec Y}$ and $Y_D$ and fall into four classes: (i) velocity transformations in the $Y_i$ directions, (ii) velocity transformation in the $Y_D$ direction, (iii) rotations in the $Y_i-Y_j$ planes, and (iv) rotations in the $Y_i-Y_D $ planes; the infinitesimal forms of these and the corresponding transformations for the conformal $\tau,{\vec x}$ coordinates are:
%\begin{widetext}
\begin{subequations}
\begin{eqnarray}
\delta Y_0=\epsilon Y_i;\ \ \delta Y_i=\epsilon Y_0 &\Longrightarrow & \delta\tau=-\epsilon\tau x_i:\ \ \delta x_i= -\epsilon [x_ix_j+
\delta_{ij}(\tau^2-1-{\vec x}\cdot{\vec x})/2]\, , \label{dSsa}\\
\delta Y_0=\epsilon Y_D;\ \ \delta Y_D=\epsilon Y_0 &\Longrightarrow & \delta\tau=-\epsilon\tau;\ \ \delta X_i=-\epsilon x_i \label{dSsb}\, ,\\
\delta Y_i=\epsilon Y_j;\ \ \delta Y_j=-\epsilon Y_i; &\Longrightarrow &\delta x_i=\epsilon x_j;\ \ \delta x_j=-\epsilon x_i\label{dSsc}\, ,\\
\delta Y_D=\epsilon Y_i;\ \ \delta Y_i=-\epsilon Y_D &\Longrightarrow & 
\delta\tau=-\epsilon\tau x_i;\ \ \delta x_j=-\epsilon [x_ix_j+\delta_{ij}(\tau^2+1-{\vec x}\cdot{\vec x})/2]\, ;\label{dSsd}\, .
\end{eqnarray}
\end{subequations}
%\end{widetext}
For any interaction on the background de Sitter space there exists an energy-momentum tensor, $\Theta_{\mu\nu}$; even though the explicit appearance of the coordinate $\tau$  spoils the conservation of this tensor, it can be used to obtain the infinitesimal generators of these transformations, $S_\nu=\delta\tau \Theta_{0\nu}-\sum_i\delta x_i \Theta_{i\nu}$, or more specifically
%\begin{widetext}
\begin{subequations}
\begin{eqnarray}
S^{(a;i)}_\nu&=&-\tau x_i\Theta_{0\nu}+\sum_j [x_ix_j+\delta_{ij}(\tau^2-1-{\vec x}\cdot{\vec x})/2]\Theta_{j\nu}\, ,\label{dSgena}\\
S^{(b)}_\nu&=&-\tau\theta_{0\nu}+\sum_jx_j\Theta_{j\nu}\, ,\label{dSgenb}\\
S^{(c;i,j)}_\nu&=&x_i\Theta_{j\nu}-x_i\Theta_{i\nu}\, ,\label{dSgenc}\\
S^{(d;i)}_\nu&=&-\tau x_i\Theta_{0\nu}+[x_ix_j+\delta_{ij}(\tau^2+1-{\vec x}\cdot{\vec x})/2]\Theta_{j\nu}\, ;\label{dSgend}
\end{eqnarray}
\end{subequations}
%\end{widetext}
{\em all indices are raised and lowered by the flat space Minkowski metric $\eta_{\mu\nu}$.}

As mentioned $\Theta_{\mu\nu}$ is not conserved, however 
\begin{equation}\label{conscond}
\tau\partial^{\alpha}\Theta_{0\alpha}=\Theta^{\alpha}_{\ \alpha}
\end{equation}
ensures the conservation of all the de Sitter currents (\ref{dSgena}--\ref{dSgend}), i.e. $\eta^{\mu\nu}\partial_\mu S^{(..)}_\nu=0$ 
As we shall note, the field equations of motion insure (\ref{conscond}); below, Sec.~(\ref{Anomaly}), we show that quantum corrections violate this relation.
\section{Two Dimensional Models}\label{2-d}
In this section we shall study this interacting fields in a (1+1) dimensional de Sitter background.  In flat (1+1) Minkowski space there are several interacting theories that can be solved exactly. Among these are: (i) the Thirring model \cite{Thirring}, (ii) massless QED \cite{Schwinger}, and (iii) spin-0 with a sine-Gordon interaction $\sim \cos(2\sqrt{\pi}\phi)\, .$  The reason these interacting field theories can be solved is that there is a correspondence \cite{{Coleman},{Kog-Suss}, {Mandelstam:1975hb},{Bander:1975pi}}  wherein spinor fields can be written in terms of spin-0 ones and for the cases cited above the interacting theory is expressible as a free theory with opposite statistics. 

Such a correspondence between bosonic and fermionic formulations can be extended to a background de Sitter space.   The two interacting fermion models, (i) and (ii) above, go over to free spin-0 theories preserving de Sitter symmetry. No instability of de Sitter space is indicated. The case of bosons interacting by a sine-Gordon term, (iii) above, corresponds to a free, massive spin-$\frac{1}{2}$ field theory albeit with a mass term that depends on the de Sitter time, thus explicitly breaking de Sitter symmetry. A further analysis of this model shows that $\ln\langle0|S|0\rangle$ has an infinite real part. indicating a vacuum instability. 
\subsection{Lagrangians  in (1+1) de Sitter Space}\label{corres}
In this two dimensional curved space with the metric (\ref{metric}) fields, propagators and Lagrangians are conformally related to the corresponding expressions in flat Minkowski space \cite{Birrell:1982ix}. For fields these conformal transformations are
\begin{eqnarray}\label{conf_trans}
 \phi_M &\leftrightarrow &\phi_{dS}\ \ \ \ \ \ \ \ \ \ \ \ {\rm spin\  0}\, ;\nonumber\\
\psi_M &\leftrightarrow&\psi_{dS}/\tau\ \ \ \ \ \ \ {\rm spin\  1/2}\, ;\\
A_{\mu;M}& \leftrightarrow & \tau^2A_{\mu;dS}\ \ \ \ \ {\rm spin\  1}\, .\nonumber
\end{eqnarray}
The metric tensors implied by (\ref{metric}) are: $g_{0,0}=-g_{1,1}=\tau^{-2}\, ,g_{0,1}=0 $ with $\sqrt{-g}=\tau^{-2}$; the corresponding {\it zweibeins}, $e^\mu_a$, which we need for a discussion of the spinor dynamics  are; $e^0_0=\tau\, ,e^1_1=\tau\, , e^0_1=e^1_0=0$. The connection tensor $\Gamma_\mu=0$. 
The action for a free, neutral, massive scalar field, $\phi$, is
\begin{equation}\label{free_lagr_0}
S_0=\frac{1}{2}\int d\tau dx \sqrt{-g}\left(g^{\mu\nu}\partial_\mu\phi\partial_\nu\phi-m_b^2\phi^2\right)=
  \frac{1}{2} \int d\tau dx \left(\partial_0\phi\partial_0\phi-\partial_1\phi\partial_1\phi-m_b^2\frac{\phi^2}{\tau^2}\right)\, ;
\end{equation}
the one for a free massive spinor  $\psi$ is
\begin{eqnarray}\label{free_lagr_1/2}
 S_{\frac{1}{2}}& =&\int d\tau dx \sqrt{-g}\Bigg[ \frac{i}{2}
\left({\bar\psi}e^{\mu,a}\gamma_a\partial_\mu\psi-e^{\mu,a}\partial_\mu{\bar\psi}\gamma_a\psi\right)-m_f{\bar\psi}\psi
\Bigg]\nonumber\\
&= &\int d\tau dx \Big[ \frac{i}{2\tau}\left({\bar\psi}\gamma_0\partial_0\psi-\partial_0{\bar\psi}\gamma_0\psi
-{\bar\psi}\gamma_1\partial_1\psi+\partial_1{\bar\psi}\gamma_1\psi\right)-m_f\frac{{\bar\psi}\psi}{\tau^2}\Big]\, ;
\end{eqnarray}
and the one for a massless vector field $A_\mu$, in the gauge $A_1=0$ 
\begin{equation}\label{free_lagr_1}
S_1=\int d\tau dx\sqrt{-g} \frac{-1}{4}F_{\mu\nu}F_{\lambda\sigma}g^{\mu\lambda}g^{\nu\sigma}=-\int
d\tau dx \frac{\tau^2}{2}(\partial_1A_0)^2\, .
\end{equation}
The conformal transformation in (\ref{conf_trans}) can be read of from the Lagrangian correspondences above.  

From (\ref{free_lagr_1/2}) we note that the momentum conjugate to $\psi$ is
\begin{equation}\label{spinor_pi}
\pi_\psi=\frac{\delta S_{\frac{1}{2}}}{\delta \partial_0\psi}=\frac{i}{\tau}\psi^\dag\, ,
\end{equation}
implying the equal-$\tau$ anticommutation relation
\begin{equation}\label{anticomm}
\left\{\psi_a(\tau,x),\psi_b^\dag(\tau,y)\right\}=\tau\delta(x-y)\delta_{ab}\, .
\end{equation}
\subsection{Fermi-bose field correspondence}
The expression for fermi fields in terms of bose ones in Ref. \cite{Bander:1975pi},  eq.(3.9), valid for Minkowski space together with (\ref{anticomm}) tells us what modification we need to make in order to obtain a similar relation valid for de Sitter space.
\begin {eqnarray}\label{sp_bos_rel}
\psi_1(\tau,x)&=&\left(\frac{\Lambda \tau}{2\pi\gamma}\right)^{1/2}\exp[-i\sqrt{\pi}\Phi_+(\tau,x)]\nonumber\\
          &{}&\\
\psi_2(\tau,x)&=&\left(\frac{\Lambda \tau}{2\pi\gamma}\right)^{1/2}\exp[-i\sqrt{\pi}\Phi_-(\tau,x)]\nonumber\, .
\end{eqnarray}
In the above $\Lambda$ is an ultra violet cut-off, $\gamma=0.577\cdots$ is the Euler-Mascheroni constant and $\Phi_{\pm}$
depends on a free massless bose field $\phi(\tau,y)$,
\begin{equation}
\Phi_{\pm}=\int_{-\infty}^x dye^{y/R}[\partial _\tau\phi(\tau,y)\pm\partial_y\phi(\tau,y)\, ;
\end{equation}
$R$ is a spatial cutoff and the limit $R\rightarrow\infty$ will be taken at the end of all calculations.  
It is the factors $\tau^{1/2}$ in front of the identities of (\ref{sp_bos_rel}) that distinguish this fermion-boson correspondence from the one in flat Minkowski space. 
\subsubsection{Composite Operators}\label{comp_oper}
Using (\ref{sp_bos_rel}) we obtain directly the translation of fermion mass operators into the language of bose fields
\begin {eqnarray}\label{mass_oper_1}
:{\bar\psi}\psi:&=&\frac{\tau\Lambda}{\pi\gamma}\cos\big[2\sqrt{\pi}\int_{-\infty}^x dy e^{y/R}\partial_y,\phi(\tau,y)\big]\, ,\nonumber\\
&{}&\\
:{\bar\psi}\gamma_5\psi:&=&i\frac{\tau\Lambda}{\pi\gamma}\sin\big[2\sqrt{\pi}\int_{-\infty}^x dy e^{y/R}\partial_y\phi(\tau,y)\big]\, .\nonumber
\end{eqnarray}
Bearing in mind the caveats expressed in Ref. \cite{Bander:1975pi}, it is convenient for comparing boson and fermion Lagrangians or actions  to set $R=\infty$ and obtain 
\begin{eqnarray}\label{mass_oper_2}
:{\bar\psi}\psi:&=&\frac{\tau\Lambda}{\pi\gamma}\cos 2\sqrt{\pi}\phi(\tau,x)\, ,\nonumber\\
  &{}&\\
:{\bar\psi}\gamma_5\psi:&=&i\frac{\tau\Lambda}{\pi\gamma}\sin 2\sqrt{\pi}\phi(\tau,x)\, .\nonumber
\end{eqnarray}
Again, it is the extra factors involving the conformal time $\tau$ that differentiate this correspondence from the one in flat space 
and it is these terms that will be responsible for breaking de Sitter symmetry for interacting theories.

We now turn to current operators. First we note that the Noether current and axial current obtained from (\ref{free_lagr_1/2}) are 
\begin{eqnarray}\label{noth-curr}
j_\mu &=&\frac{1}{\tau}:{\bar\psi}\gamma_\mu\psi:\, ,\nonumber\\
&{}&\\
j^5_\mu &=&\frac{1}{\tau}:{\bar\psi}\gamma_\mu\gamma_5\psi:\, .\nonumber
\end{eqnarray}
This time the extra factors involving $\tau$ cancel and the correspondence is as in flat space.
\begin{eqnarray}\label{curr_oper}
j_\mu(\tau,x)&=&\frac{\epsilon_{\mu\nu}}{\sqrt{\pi}}\partial^\nu\phi(\tau,x)\, ;\nonumber\\
&{}&\\
j_\mu^5(\tau,x)&=&\frac{1}{\sqrt{\pi}}\partial_\mu\phi(\tau,x)\, .\nonumber
\end{eqnarray}

\subsection{Interacting Theories -- Correspondence}\label{interact}
We shall look at a class of two dimensional theories that, in one language, bose or fermi, have non-trivial interactions, while in the other language are free field theories.  These are: (i) the Thirring Model, (ii) massless fermion QED and (iii) a sine-Gordon interaction. 
\subsection{Massless Thirring model $\leftrightarrow$ Free massive boson}\label{thirring}The action for a fermion with a current-current interaction, Thirring model, on a de Sitter space is
\begin{equation}\label{thirr_action_fermi}
S_{\rm Thirring}=\int d\tau dx \Big[ \frac{i}{2\tau}\left({\bar\psi}\gamma_0\partial_0\psi-\partial_0{\bar\psi}\gamma_0\psi
-{\bar\psi}\gamma_1\partial_1\psi+\partial_1{\bar\psi}\gamma_1\psi\right)-\frac{g}{2}(j_0j_0-j_xj_x\Big]\, ,
\end{equation}
which, using (\ref{curr_oper}), is equivalent to a free mass-less bose action with the fermi field--bose field identification (\ref{sp_bos_rel}) rescaled to
\begin{equation}
\psi_{1,2}=\left(\frac{\tau\Lambda}{\pi\gamma}\right)^{1/2}\exp\left\{-i\sqrt{\pi}\int_{-\infty}^x  dy e^{y/R}\left[
   \partial_0\phi/\beta \pm\beta\partial_y\phi\right]\right\}\, ,
\end{equation}
and $\beta=(1+g/\sqrt{\pi})$. De Sitter symmetry holds in both formulations. 
\subsubsection{Massless QED}\label{qed}
With the photon field in the $A_1=0$ gauge, the fermi action is 
\begin{eqnarray}\label{qed_action}
S_{\rm QED}&=&\int d\tau dx \left[ \frac{i}{2\tau}\left({\bar\psi}\gamma_0\partial_0\psi-\partial_0{\bar\psi}\gamma_0\psi
-{\bar\psi}\gamma_1\partial_1\psi+\partial_1{\bar\psi}\gamma_1\psi\right)-ej_0A_0+\frac{\tau^2}{2}\left(\partial_1A_0\right)^2\right]\, . \nonumber\\
&{}&
\end{eqnarray}
Solving the equation of motion for $A_0$ an using (\ref{curr_oper}) results in a scalar field action as in (\ref{free_lagr_0}) with $m^2_b=e^2/\pi$.  Again, the de Sitter symmetry is valid in both formulations. 

\subsubsection{Sine-Gordon Interaction}\label{Sine_Gordon_sect}
We consider a $\cos\beta\phi$ interaction with a special value for $\beta$, namely $\beta=2\sqrt{\pi}$. 
\begin{equation}\label{sine_gordon}
S_{\rm sine-Gordon}=
  \frac{1}{2} \int d\tau dx \left[\partial_0\phi\partial_0\phi-\partial_1\phi\partial_1\phi-\frac{g}{\tau^2}\cos\left(2\sqrt{\pi}\phi\right)\right]\, .
\end{equation}
Eq. (\ref{mass_oper_2}) allows us to identify the above with $S_{\frac{1}{2}}$ of ( \ref{free_lagr_1/2}) with $m_f=g\pi\gamma/(\tau\Lambda)$\, .
This explicit $1/\tau$ behavior of the fermion mass breaks de Sitter symmetry. Below we shall look at this case in greater detail.

In the fermionic language the action is 
\begin{equation}\label{taumass}
S_{\tau-{\rm dep-mass}}\int d\tau dx \frac{i}{\tau}\Big({\bar\psi}\gamma_0\partial_0\psi-{\bar\psi}\gamma_1\partial_1\psi
+\frac{1}{2\tau}{\bar\psi}\gamma_0\psi\Big)-M\frac{{\bar\psi}\psi}{\tau^3}\, 
\end{equation}
with $M$ related to the strength of the sine-Gordon interaction.   
The vacuum to vacuum amplitude is 
\begin{equation}\label{vac-vac-1}
\langle 0, {\rm out}|0, {\rm in}\rangle=\exp {\rm tr}  \ln\left(i\gamma^\mu\partial_\mu-M/\tau^3\right)\, ;
\end{equation}
to evaluate the above we need the eigenvalues of the Dirac operator, with a non constant mass term $\gamma^\mu{\partial}_\mu-M/\tau^3$. If $\psi$ is an eigenfunction of this operator then $\gamma_5\psi$ is an eigenfunction of $-\gamma^\mu{\partial}_\mu-M/\tau^3$ with the same eigenvalue and we may replace (\ref{vac-vac-1}) with
\begin{equation}\label{vac-vac-2}
\langle 0, {\rm out}|0, {\rm in}\rangle=\exp {\frac{1}{2}\rm tr}  \ln\left(i\gamma^\mu{\partial}_\mu-M/\tau^3\right)\left(-i\gamma^\mu{\partial}_\mu-M/\tau^3\right)\, ,
\end{equation}
which requires us to look at the eigenvalues of $\left(i\gamma^\mu{\partial}_\mu-M/\tau^3\right)\left(-i\gamma^\mu{\partial}_\mu-M/\tau^3\right)=
\partial^2+M^2/\tau^6+3i\gamma_0M/\tau^4\, .$ After rotating to Euclidian time, $\tau\rightarrow it_{\rm E}$ we want to determine the reality properties of the eigenvalue of the operator (with $e^{ikx}$ spatial dependence and diagonal $\gamma_0$); 
\begin{equation}\label{euclid_oper}
-\partial^2_{t_{\rm E}}+k^2-M^2/t_{\rm E}^6\pm 3iM/t^4_{\rm E})\, .
\end{equation}
Aside from the explicit imaginary terms, the real part of the above operator is just a one dimensional Schr\"{o}dinger equation with an $1/r^6$ attractive potential resulting in an infinite number  of negative eigenvalues whose logarithms have imaginary parts. The trace in (\ref{vac-vac-1}), after rotating to Euclidian time, introduces an other factor of $i$ resulting in an infinite sum of real contributions to the exponent in (\ref{vac-vac-1}) and a vanishing $\langle 0, {\rm out}|0, {\rm in}\rangle$ amplitude. 

\section{Anomalies}\label{Anomaly}
The discussion in the previous section is not totally satisfactory for two reasons. It is limited to (1+1) dimensions and one should be able to discover the breaking of de Sitter symmetry directly in the boson sector, without recourse to a boson-fermion equivalence.  We shall remedy both in this section. 

Earlier we showed that the validity of eq.~(\ref{conscond}) determines the conservation of currents generating de Sitter transformations. We shall examine whether in a theory of an interacting scalar field, $\phi(\tau,{\vec x})$ governed by the action 
\begin{equation}\label{action}
{\cal S}=\int d\tau d^dx \tau^{-D}\left[\frac{\tau^2}{2}\partial^\mu\phi\phi\partial_\mu\phi-V(\phi)\right] \, ,
\end{equation}
this condition is satisfied when quantum loop corrections are included.  Details of calculations will be presented for the case where 
\begin{equation}\label{potential}
V(\phi)=\frac{m^2}{2}\phi^2+\frac{g}{4!}\phi^4\, 
\end{equation}
and then generalized to arbitrary $V(\phi)$.
Using equations of motion obtained from (\ref{action}) it is straight forward to show that the corresponding energy-momentum tensor
\begin{equation}\label{en-mom-tensor}
\Theta_{\mu\nu}=\tau^{2-D}\partial_\mu\phi\partial_\nu\phi-\eta_{\mu\nu}\left[\tau^{2-D}\frac{1}{2}
\partial_\alpha\phi\partial^\alpha\phi-\tau^{-D}V(\phi)\right]\, 
\end{equation}
does satisfy (\ref{conscond}). We shall show, however, that to order $g$ the {\em regularized} one loop correction does not satisfy this relation. This will be true for the propagator advocated in \cite{Polyakov:2007mm, Polyakov:2009nq} as well as  all other $\alpha$-vacua propagators save the Euclidian one. Explicitly, the propagator we shall use is 
\begin{equation}\label{polyakovprop}
D_m(x_1,x_2)=C(1-z_{12}^2)^{-(D-2)/4}{\cal Q}_{-\frac{1}{2}+i\nu(m)}^{(D-2)/2}(z_{12})\, ;
\end{equation}
with $Y_i$ the coordinates in the embedding space corresponding to the point $x_i$ in the de Sitter space, $ z_{ij}=Y_i\cdot Y_j$. $\cal Q$ is an associated Legendre function of the second kind and $\nu(m)=\sqrt{m^2-(D-1)^2/4}$;  the constant $C$ depends only on the dimension of the de Sitter space and is chosen to insure a correct residue at $z_{12}=1$, corresponding to $Y_1=Y_2$. In addition to the ``ultraviolet" singularity at $Y_1=Y_2$ (\ref{polyakovprop}) has an ``infrared", singularity at $z_{12}=-1$, namely $Y_2=-Y_1$ or $x_2={\bar x}_1$, the point antipodal to $x_1$ \cite{Spradlin:2001pw}.  It is this singularity that will be responsible for the non-conservation of de Sitter currents.

To determine the conservation, or lack thereof,  we shall study the matrix element
\begin{equation}\label{matelem}
T_{\mu\nu}(x; y_1, y_2)=\langle T[\Theta_{\mu\nu}(x)\phi(y_1)\phi(y_2)]\rangle\, ,
\end{equation}
where the symbol $T$ in the matrix element above indicates the conformal time, $\tau$, ordered product. To zeroth order in $g$ we find
\begin{eqnarray}\label{0-order}
T_{\mu\nu}(x; y_1, y_2)&=&\left[\tau^{2-D}\partial_\mu D_m(x,y_1)\partial_\nu D_m(x,y_2)+(y_1\leftrightarrow y_2)\right]\nonumber\\
&-&\eta_{\mu\nu}\left[\tau^{2-D}\eta_{\alpha\beta}\partial_\alpha D_m(x,y_1)\partial_\beta D_m(x,y_2)-\tau^{-D}m^2D_m(x,y_1)D_m(x,y_2)\right]\, ,
\end{eqnarray}
where $\tau$ is the time associated with the $x$ coordinate. 
Up to terms involving equal time commutators, $[\Theta_{0\nu}(\tau,{\vec x}),\phi(\tau, {\vec y}_i)]$ coming from
differentiating the time ordering,  $T_{\mu\nu}(x; y_1, y_2)$ satisfies an equation analogous to (\ref{conscond}).

Aside from mass renormalizations, to order $g$ the correction to $T_{\mu\nu}(x; y_1, y_2)$ is given by Fig.~1,
 \begin{figure}[ht]
 \includegraphics[width=.5\textwidth]{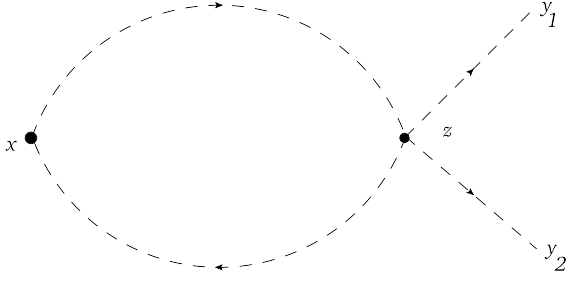}
 \caption{Lowest order loop correction to $T_{\mu\nu}(x;y_1,y_2)$.}
\label{fig1}
\end{figure}      
resulting in
\begin{eqnarray}\label{correction}
\delta T_{\mu\nu}(x; y_1, y_2)&=&g\int d^5z\delta(z^2+1) \{\tau^{2-D}\partial_\mu D_m(x,z)\partial_\nu D_m(x,z)\nonumber\\&-&\frac{\eta_{\mu\nu}}{2}[\tau^{2-D}\eta_{\alpha\beta}\partial_\alpha D_m(x,z)\partial_\beta D_m(x,z)-\tau^{-D}m^2D_m(x,z)^2]\}D_m(z,y_1)D_m(z,y_2)\, .\nonumber\\
\end{eqnarray}

For the de Sitter currents to be conserved we require that
\begin{equation}\label{conscond-2}
\Delta(x;y_1,y_2)=\tau\eta^{\mu\nu}\partial_\mu\delta T_{0\nu}(x; y_1, y_2)-\eta^{\mu\nu}\delta T_{\mu\nu}(x; y_1, y_2)=0\, .
\end{equation} 
The validity of 
\begin{equation}\label{formalcond}
\int d^5z\delta(z^2+1) \{\tau^{2-D}\partial^{\mu}\partial_\mu D(x,z)\partial_\nu D(x,z)\-\tau^{-D}m^2\partial_\nu D(x,z)^2]\}D(z,y_1)D(z,y_2)=0\, ,
\end{equation}
which follows from the equations of motion, would insure (\ref{conscond-2})
Although this relation is {\em formally} satisfied it involves products of functions singular, both, at $z=x$ and at $z={\bar x}$;  thus before we conclude anything the integral in (\ref{conscond-2}) must be regulated. As the singularity at $z=x$ is a short distance one, the curvature of the underlying space does not come into play and it is removed by the usual ultraviolet renormalization. The singularity at $z={\bar x}$ is new and requires its own regularization. 

As the residue of the pole at $z_{12}=-1$ in (\ref{polyakovprop}) does depend on the mass \cite{Erdelyi}, the regularization we use consists of {\em subtracting} from (\ref{correction})  an expression in with all propagators $D_m(x,z)=(1-z_{12}^2)^{-(D-2)/4}{\cal Q}_{-\frac{1}{2}+i\nu(m)}^{(D-2)/2}(z_{12})$ replaced by 
\[
D_M(x,z)=\left[{\cos(i\nu(m)+(D-2)/2)}/{\cos(i\nu(M)+(D-2)/2}\right](1-z_{12}^2)^{-(D-2)/4}{\cal Q}_{-\frac{1}{2}+i\nu(M)}^{(D-2)/2}(z_{12})
\]
 (the prefactor involving the cosines makes the residues at $z=-1$ in $D_m$ and $D_M$ equal) and at the end letting $M\rightarrow\infty$. The substitution, $m\rightarrow M$ is performed only in the propagators and not in the $m$ that appears explicitly in (\ref{correction}). The formal manipulations may now be carried out resulting in
\begin{equation}\label{anomaly-1}
\Delta(x;y_1,y_2)=\int d^5z\delta(z^2+1)\tau^{1-D}(M^2-m^2)D_M(x,z)^2D_m(z,y_1)D_m(z,y_2)\, .
\end{equation}
The conservation of the de Sitter currents depends on whether $\Delta\rightarrow 0$ as $M\rightarrow\infty$.  To perform the indicated integration we follow the procedure of \cite{Polyakov:2009nq}.  We shall show that in the large $M$ limit the integrand will be peaked at $z={\bar x}$ and we can replace the propagators $D_m(z,y_i)$ by $D_m({\bar x}, y_i)$. As the resultant integral is invariant under Lorentz transformations in the imbedding space, we may set $x=(0,1,0,\cdots,0)$ and as $D_M(x,z)=D_M(x\cdot z-i\epsilon)$ (the dot product being taken in the imbedding space) (\ref{anomaly-1}) becomes 9after integrating over $z_2,\cdots,z_D$)
\begin{equation}\label{anomaly-2}
\Delta(x;,y_1,y_2)\sim M^2\int dz_0dz_1 (z_0^2-z_1^2+1)_+^{(D-3)/2}D_M(z_1-i\epsilon)^2D_m({\bar x},y_1)D_m({\bar x},y_2)\, ;
\end{equation}
the $+$ subscript in $(\cdots)_+$ denotes that further integrations are to be restricted to the region where the expression inside the parenthesis is positive. The $z_0$ integration is divergent and depends on the large $z_0$ cut-off; however,  as pointed out in \cite{Polyakov:2009nq}, due to analiticity of the propagator in the lower half plane the coefficient of this cut-off is zero. The result of this integration is
\begin{eqnarray}\label{anomaly-3.5}
\Delta(x;,y_1,y_2)&\sim& M^2\int_{-1}^1 dz_1\left\{\begin{array}{cc}
                                                                            -2i\pi\epsilon(z_1)\ln(z_1^2-1) & D=2\\
                                                                               2(1-i\epsilon(z_1))\sqrt{(z_1^2-1} & D=3\\
                                                                                2(z_1^2-1)\ln(z_1^2-1) & D=4
                                                                                               \end{array}\right\}\nonumber\\
&\times&D_M(z_1-i\epsilon)^2D_m({\bar x},y_1)D_m({\bar x},y_2)\, .
\end{eqnarray}
Relying on the analiticity of the propagators in the lower half plane, we find that when the integration over $z_1$ extended over the whole real line the result is zero. Taking the cut of the logarithms and the square root in (\ref{anomaly-3.5}) to be in the interval $-1<z_1<1$
we obtain the result as an integration only over the range $-1<z_1<1$.
\begin{eqnarray}\label{anomaly-4}
\Delta(x;,y_1,y_2)&\sim& M^2\int_{-1}^1 dz_1\left\{\begin{array}{cc}
                                                                                                      -2i\pi\epsilon(z_1) & D=2\\
                                                                                                       2(1-i\epsilon(z_1))\sqrt{(1-z_1^2} & D=3\\
                                                                                                       2i\pi\epsilon(z_1)(z_1^2-1) & D=4
                                                                                               \end{array}\right\}\nonumber\\
&\times&D_M(z_1-i\epsilon)^2D_m({\bar x},y_1)D_m({\bar x},y_2)\, .
\end{eqnarray}

Using explicitly the propagator in (\ref{polyakovprop}),  $D_M(z)=(1-z^2)^{-(D-2)/4}{\cal Q}_{-\frac{1}{2}+i\nu(M)}^{(D-2)/2}(z-i\epsilon)$  we are asked to look at the large $M$ limit of $\left[\cos(i\nu(M)+(d-2)/2\right]D_M(z-i\epsilon)^2$. From \cite{Erdelyi} we find
\begin{equation}\label{Q-limit}
\cos[i\nu(M)+(D-2)/2]^{-1}D_M(z)\rightarrow 
\sqrt{\frac{\pi}{2}}e^{-M\pi}M^{(D-3)/2}\left[\frac{z+(z^2-1)^{\frac{1}{2}}]^{i\nu(M)+\frac{1}{2}}}{(z^2-1)^{\frac{1}{4}}}\right]\, ;
\end{equation}
with all $z$'s having a small negative imaginary part.
At $z=-1$ this limit is infinite while for all $z>-1$ it is zero, justifying our earlier replacement in $D_m(z,y_i)$ of $z$ by $\bar x$. The integral of the square of 
(\ref{Q-limit}) multiplied by the dimension dependent factors in (\ref{anomaly-4}) behaves as $M^{-2}$ resulting in 
\begin{equation}\label{anaomaly-5}
\Delta(x;,y_1,y_2)\sim D_m({\bar x},y_1)D_m({\bar x},y_2)\, ,
\end{equation}
or, going back to eqs.~(\ref{dSgena}--\ref{dSgend})
\begin{equation}\label{result-1}
\eta^{\mu\nu}\partial_{\mu}S^{(\cdot,i)}_\nu(x)\sim\left(\tau\frac{\partial}{\partial\tau}+x_i\frac{\partial}{\partial x_i}\right) g\phi({\bar x})^2\, .
\end{equation}
This can be generalized to any interaction of scalar fields as, eq.(\ref{action}),
\begin{equation}\label{result-2}
\eta^{\mu\nu}\partial_{\mu}S^{(\cdot,\cdot)}_\nu(x)\sim\left(\tau\frac{\partial}{\partial\tau}+x_i\frac{\partial}{\partial x_i}\right) \frac{\partial^2}{\partial\phi^2}V(\phi({\bar x})\, .
\end{equation}
As the propagator for the Euclidian vacuum has no antipodal singularity, these anomalies do not apply for that case. 

\section{Summary}\label{conclusions}
Several questions remain unanswered. The technical ones are: 
\begin{enumerate}
\item[(i)]
Are these results dependent on the infrared regularization scheme?
\item[(ii)]
 How do higher order corrections affect these results?
 \item[(iii)]
 Do interacting fermions induce a similar anomaly? The results in \cite{Bander:2010pn} would indicate that the answer is no. This is not surprising as spin-0 fields are more pathological in the infrared than 
spin-$\frac{1}{2}$ ones.  
\end{enumerate} 
A more fundamental question is: is the propagator in \cite{Polyakov:2007mm} the one to use in perturbative calculations on this positive curvature space or does the result presented here serve as another nail in the coffin of the $\alpha$-vacua \cite{Einhorn-Larsen}\cite{Banks:2002nv}?

\acknowledgments
I wish to thank Dr.~ E.~Rabinovici, Dr.~A.~Rajaraman and and Dr.~ A.~Schwimmer for discussions and suggestions. Also, I would like to thank the organizers of the H\'{e}ctor Rubinstein Memorial Symposium for the invitation to participate.

\end{document}